\documentclass[10pt]{article}
\usepackage{amsmath}
\usepackage{amssymb}
\usepackage{natbib}
\usepackage{graphicx}

\begin{document}
\bibliographystyle{plainnat}
\setcitestyle{numbers,square}

\title{\small{CRITICAL ROTATION OF GENERAL-RELATIVISTIC POLYTROPIC MODELS \\
              SIMULATING NEUTRON STARS: A POST-NEWTONIAN                  \\
              HYBRID APPROXIMATIVE SCHEME}}
\author{
Vassilis S. Geroyannis$^1$, Vasileios G. Karageorgopoulos$^2$ \\
\textit{$^{1,2}$Department of Physics, University of Patras, Patras, Greece} \\ 
\textit{$^1$vgeroyan@upatras.gr, $^2$vkarageo@upatras.gr}}

\maketitle

\begin{abstract}
We develop a ``hybrid approximative scheme'' in the framework of the post-Newtonian approximation for computing general-relativistic polytropic models simulating neutron stars in critical rigid rotation. We treat the differential equations governing such a model as a ``complex initial value problem'', and we solve it by using the so-called ``complex-plane strategy''. We incorporate into the computations the complete solution for the relativistic effects, this issue representing a significant improvement with regard to the classical post-Newtonian approximation, as verified by extended comparisons of the numerical results.  
\\
\\
\textit{Keywords:} critical rotation; general-relativistic polytropic models; hybrid approximative scheme; neutron stars; post-Newtonian approximation
\end{abstract}

{\section{Introduction}
\label{intro}
The original contributions to the study of rapidly rotating neutron stars in the framework of the ``post-Newtonian approximation'' (PNA) are due to Chandrasekhar \citep{C65a}, Krefetz \citep{K67}, and Fahlman \& Anand \citep{FA71}. The problem of fast rigid rotation of  neutron stars in hydrostatic equilibrium is treated in \citep{FA71} by considering the relativistic and rotational effects acting on a nonrotating Newtonian configuration obeying the polytropic ``equation of state'' (EOS, EOSs). However, there are certain reasons leading the PNA of first-order in the gravitation parameter $\sigma$ to failure when $\sigma \geq 0.01$. A discussion on this matter can be found in \citep{T65} (Appendix). A further discussion (\citep{FA71}, Sec.~5) verifies the negative conclusions of \citep{T65} and focuses on the imposed limitations when  applying this PNA's scheme to several astrophysical objects, since, unfortunately, values of interest lie in the vicinity of $\sigma \simeq 0.1$.

In a recent study \citep{GK14}, we revisit the problem by assuming the relativistic and rotational effects as decoupled perturbations, and by applying to PNA the so-called ``complex plane strategy'' (CPS). This method consists in solving all differential equations involved in the PNA's computational scheme in the complex plane. Numerical integrations are resolved by the Fortran code \texttt{DCRKF54} \citep{GV12}, which is a Runge--Kutta--Fehlberg code of fourth and fifth order, modified so that to integrate ``initial value problems'' (IVP, IVPs) established on systems of first-order ``ordinary differential equations'' (ODE, ODEs) of complex-valued functions in one complex variable along prescribed complex paths.

As discussed in \citep{GK14} (Sec.~5.2), CPS could proceed independently of the particular perturbation approach used. For instance, CPS could be applied to a PNA's scheme of up to second order in $\sigma$, as developed in \citep{CN69}. But, most interesting, CPS could cooperate with a ``hybrid approximative scheme'' (HAS) of PNA (\citep{GK14}, Sec.~5.2), in which the complete solution of the relativistic distortion, as developed in \citep{T65}, is involved. 
In this study, we extend the numerical experiments started in \citep{GK14} (Sec.~5.2), by applying HAS to general-relativistic polytropic models of critically rotating neutron stars with $\sigma$ up to $\simeq 0.3$. 

We do not intend to repeat here extended parts of \citep{GK14}, except for certain significant issues. For clarity and convenience, we use the same conventions, definitions, and symbols with those in \citep{GK14}.

\section{The Hybrid Approximative Scheme}
\subsection{Preliminaries}
\label{preliminaries}
In this study, we assume that the pressure $p$ and the rest-mass density $\rho$ obey the polytropic ``equation of state'' (EOS)  
\begin{equation}
p = K \, \rho^\Gamma = 
K \, \rho^{1+\left( 1/n \right)}, 
\label{2.5}
\end{equation}
where $K$ is the polytropic constant, $\Gamma$ the adiabatic index defined by $\Gamma = 1 + (1/n)$, $n$ the polytropic index, and the normalization equations for the rest-mass density $\rho$ and the coordinate $r$ are defined by
\begin{equation}
\rho = \rho_{\mathrm{c}} \, \Theta^n, \qquad
r = \left[ \frac{(n+1) \, p_{\mathrm{c}}}{4 \, \pi \,  
     G \, \rho_{\mathrm{c}}^2} \right]^{1/2} \xi 
=   \left[ \frac{(n+1) \, K \, \rho_\mathrm{c}^{\left(1/n\right)}}{4 \, \pi \,  
     G \, \rho_\mathrm{c}} \right]^{1/2} \xi
=   \alpha \, \xi,
\label{2.5bc}
\end{equation}
where $\rho_\mathrm{c}$ is the central density, $\Theta(\xi, \, \mu)$ (with $\mu = \cos(\vartheta)$) the Lane--Emden function, $p_\mathrm{c}$ the central pressure, and $G$ the gravitation constant. The central density $\rho_\mathrm{c}$ is chosen to be the density unit in the so-called ``classical polytropic units'' (cpu), and the model parameter $\alpha$ is chosen to be the length unit in cpu; accordingly, $\Theta^n$ is the cpu measure of the rest-mass density $\rho$ and $\xi$ the cpu measure of the coordinate $r$.

The ``rotation parameter'' $\upsilon$, representing the effects of rotation, and the ``gravitation parameter'' $\sigma$ (also called ``relativity parameter''), representing the post-Newtonian effects of gravitation, are then defined by (\citep{GK14}, Eqs. (7a) and (7b), respectively)
\begin{equation}
\upsilon = \frac{\Omega^2}{2 \, \pi \, G \, \rho_{\mathrm{c}}}, \qquad 
\sigma = \frac{1}{c^2} \, \, \frac{p_{\mathrm{c}}}{\rho_{\mathrm{c}}}.  
\label{2.7}
\end{equation}

In the framework of PNA, the function $\Theta(\xi,\mu)$ can be expressed as (\citep{GK14}, Eq.~(9)) 
\begin{equation}
\begin{aligned}
\Theta(\xi,\mu) &= \sum_{i=0,\,2}^4 P_i(\mu) \, \Theta_i(\xi)   \\
&= \alpha_0 \, \theta_{00}(\xi) \, P_0(\mu)  \\
&+ \, \alpha_1 \left[ \theta_{10}(\xi) \, P_0(\mu) +  
                    A_{12} \theta_{12}(\xi) P_2(\mu) \right] \\
&+ \, \alpha_2 \left\{ \theta_{20}(\xi) \, P_0(\mu) + \left[ \theta_{22}(\xi) + A_{22} \theta_{12}(\xi) \right] P_2(\mu) \right. \\ 
&+ \left. \qquad \qquad \qquad \qquad \, \left[ \theta_{24}(\xi) + A_{24} \theta_{14}(\xi) \right] P_4(\mu) \right\} \\
&+ \, \alpha_3 \, \theta_{30}(\xi) \, P_0(\mu),
\label{2.9a} 
\end{aligned}
\end{equation}
where $\alpha_i$ are the perturbation parameters (\citep{FA71}, Eq.~(24)): $\alpha_0=1$, $\alpha_1=\upsilon$, $\alpha_2=\upsilon^2$, and $\alpha_3=\sigma$.
The functions $\theta_{ij}$ are involved in the differential equations (\citep{GK14}, Eq.~(12)) 
\begin{equation}
\frac{d^2\theta_{ij}}{d\xi^2} + \frac{2}{\xi} \, \frac{\theta_{ij}}{d\xi} - \frac{j(j+1)}{\xi^2} \, \theta_{ij}  =S_{ij} 
\label{2.10} 
\end{equation}
with $i = 0, \, 1, \, 2, \, 3,$ and $j = 0, \, 2, \, 4$, solved in view of the initial conditions~(26) of \citep{GK14}. The parameters $A_{ij}$ (\citep{GK14}, Eqs.~(24)--(25)) multiply properly the  homogeneous solutions of $\theta_{ij}$ (\citep{FA71}, Eqs.~(42) and (43)), so that the boundary conditions~(16) of \citep{GK14} be satisfied.
The functions $S_{ij}$ are given by Eq.~(13) of \citep{GK14}.

\subsection{The numerical method}
We now consider HAS as a computational scheme applied on PNA of \citep{GK14}, in which the relativistic distortion participates with its complete solution, as it has been developed and computed in \citep{T65}. By substituting the complete solution $\Theta_\sigma$ for the relativistic effects in the place of the sum $\alpha_0 \, \theta_{00}(\xi) + \alpha_3 \, \theta_{30}(\xi)$ (\citep{GK14}, Eq.~(57)), we obtain the form 
\begin{equation}
\begin{aligned}
\Theta(\xi,\mu) &= 
\Theta_\sigma \, P_0(\mu) \\
&+ \,\alpha_1 \left[ \theta_{10}(\xi) \, P_0(\mu) +  
                    A_{12} \theta_{12}(\xi) P_2(\mu) \right] \\
&+ \, \alpha_2 \left\{ \theta_{20}(\xi) \, P_0(\mu) + \left[ \theta_{22}(\xi) + A_{22} \theta_{12}(\xi) \right] P_2(\mu) \right. \\ 
&+ \left. \qquad \qquad \qquad \qquad \, \left[ \theta_{24}(\xi) + A_{24} \theta_{14}(\xi) \right] P_4(\mu) \right\}. 
\label{2.9b} 
\end{aligned}
\end{equation}

To compute the function $\Theta_\sigma$, we use the Oppenheimer--Volkoff equations of hydrostatic equilibrium (cf.~\citep{T65}, Eqs.~(19) and (20)),
\begin{equation}
\frac{d\Theta_\sigma}{d\xi} = 
- \, \frac{1}{\xi^2} \left( \Upsilon_\sigma + \sigma \, \xi^3 \, \Theta_\sigma^{n+1} \right)   \, \frac{\left[ 1 + (n+1) \, \sigma \, \Theta_\sigma \right]}{1 - 2 \, \sigma \, (n+1) \,  \Upsilon_\sigma/\xi},
\label{T6501}
\end{equation}
\begin{equation}
\Upsilon_\sigma' = \xi^2 \, \Theta_\sigma^n \left( 1 + \sigma \, n \, \Theta_\sigma \right),
\label{T6502}
\end{equation}
where the function $\Upsilon_\sigma$ is defined by (cf. \citep{T65}, Eq.~(18))
\begin{equation}
m(r) = 4 \, \pi \, \alpha^3 \, \rho_\mathrm{c} \Upsilon_\sigma(\xi);
\end{equation}                           
$m(r)$ is the total mass interior to a sphere of radius $r$ (cf. \citep{T65}, Eq.~(12)). In the Newtonian limit $\sigma=0$, Eqs.~(\ref{T6501}) and (\ref{T6502}) reduce to the classical Lane--Emden equation (Eq.~(\ref{2.10}) with $i=j=0$). In the relativistic case $\sigma > 0$, $\Theta_\sigma$ is the total distortion owing to relativistic effects and can be written as (\citep{GK14}, Eq.~(57))
\begin{equation}
\Theta_\sigma = \theta_{00} + \sum_{i=1}^{\infty} \sigma^i \, \theta_{3(i-1)}.
\label{theta3}
\end{equation}
The PNA's scheme in \citep{GK14} includes terms of first order in $\sigma$; in this case, the sum in Eq.~(\ref{theta3}) contains the single term $\sigma \, \theta_{30}$. When with infinite terms, the sum in Eq.~(\ref{theta3}) becomes equal to $\Theta_\sigma - \theta_{00}$. The computational basis of HAS consists in using the complete solution in the relativistic distortion and perturbation terms of up to second order in $\upsilon$ with respect to the rotational distortion.      

The initial conditions for solving the differential equations~\eqref{2.10}, ~\eqref{T6501}, and \eqref{T6502} are written as (cf.~\citep{GK14}, Eqs.~(26)) 
\begin{equation}
\begin{aligned}
&\theta_{00}=1, \qquad \, \, 
\frac{d\theta_{00}}{d\xi}=0, \qquad &\mathrm{at} \, \, \, \xi=0, \\
&\theta_{ij}=0, \qquad \, \, \,  
\frac{d\theta_{ij}}{d\xi}=0, \, \, i = 1,\,2, \, \, j = 0, 
 \qquad &\mathrm{at} \, \, \, \xi=0, \\
&\theta_{ij}=\xi^j, \qquad \frac{d\theta_{ij}}{d\xi}=j \, \xi^{j-1}, \, \, i = 1,\,2, \, \, j \geq 2, \qquad &\xi \in \delta(0), 
\label{2.18}
\end{aligned} 
\end{equation}
where the interval $\delta(0)$ lies in the vicinity of zero, and
\begin{equation}
\Theta_\sigma = 1, \qquad \Upsilon_\sigma = 0.
\label{2.18TY}
\end{equation}

\subsection{The complex-plane strategy} 
\label{theCPS}
Equation~(\ref{2.10}) yields for $i=j=0$ the classical Lane--Emden equation, which, integrated along a prescribed interval 
$\mathbb{I}_\xi = [\xi_\mathrm{start} = 0, \,\, \xi_\mathrm{end}] \subset \mathbb{R}$
with initial conditions~(\ref{2.18}a,\,b)
gives the Lane--Emden function $\theta_{00}[\mathbb{I}_\xi \subset \mathbb{R}] \subset \mathbb{R}$. To avoid the indeterminate form $\theta_{00}'/\xi$ at the origin, we start integration at a point $\xi_\mathrm{start} = \xi_0$ close to the origin. Since $\xi_0$ is small, the initial conditions~(\ref{2.18}a,\,b) are valid at the starting point $\xi_0$ as well. So, the integration interval becomes 
$\mathbb{I}_{\xi0} = [\xi_0,\,\xi_\mathrm{end}] \subset \mathbb{R}$.

The Lane--Emden function $\theta_{00}$ becomes zero at its first root $\Xi_1$,  $\theta_{00}(\Xi_1) = 0$. Beyond the first root $\Xi_1$, $\xi > \Xi_1$, $\theta_{00}$ changes sign, $\theta_{00}(\xi) < 0$. Accordingly,  $\theta_{00}^n$ is undefined beyond $\Xi_1$, since raising a negative real number to a real power is not defined in $\mathbb{R}$. To remove this syndrome, we can define $\theta_{00}$ as a complex-valued function in one real variable $\xi$ with $\xi \in \mathbb{I}_{\xi0}$, 
$\theta_{00}[\mathbb{I}_{\xi0} \subset \mathbb{R}] \subset \mathbb{C}.$

 Since $n \in \mathbb{R}$, the term $\theta_{00}^n$ suffers from a ``non-monodromy syndrome'' due to the fact that multiple-valued logarithmic functions are involved in the representation of $\theta_{00}^n$ (see e.g. \citep{Chu60}, Secs.~26--28). To remove this syndrome, we proceed by defining an ``auxiliary Lane--Emden function'' $\chi$ such that $\theta_{00} = \chi^N$ (\citep{GK14}, Eq.~(35)), 
where the involved integer $N$ is chosen so that the term $\theta_{00}^n = \chi^{Nn}$ be transformed into a ``raised-to-integer-power'' term. 
The ``modified Lane--Emden equation'' for $\chi$ with its initial conditions (\citep{GK14}, (Eqs.~(36) and (37), respectively) can be transformed into an equivalent IVP in two first-order ODEs (\citep{GK14}, (Eqs.~(38) and (39))
\begin{equation}
\frac{d\chi}{d\xi} = \phi,
\label{MLEEe1}
\end{equation}
\begin{equation}
\frac{d\phi}{d\xi} = - \, \frac{2}{\xi} \, \phi - \frac{N-1}{\chi} \, \phi^2 -  
                          \frac{1}{N} \,\, \chi^{N \left(n-1\right)+1}, 
\label{MLEEe2}
\end{equation}
where $\chi[\mathbb{I}_{\xi0} \subset \mathbb{R}] \subset \mathbb{C}$ and $\phi[\mathbb{I}_{\xi0} \subset \mathbb{R}] \subset \mathbb{C}$, which are solved with initial conditions
\begin{equation}
\chi(\xi_0) = \theta_{00}(\xi_0)^{1/N}, \qquad \phi(\xi_0) = 0.
\label{mivs2}
\end{equation}

To avoid a further singularity at $\Xi_1$, owing to the term $\phi = \chi'$, we assume that 
the independent variable $\xi$ is a ``complex distance'', $\xi \in \mathbb{C}$, and that the integration proceeds along a prescribed complex path parallel to the real axis and at a relatively small imaginary distance from it, playing the role of a complex detour. This alternative consists in performing numerical integration along a contour $\mathfrak{C} \subset \mathbb{C}$, being parallel to the real axis $\mathbb{R}$ and distancing $i\,\breve{\xi}_0$ from it, i.e. along the straight line-segment
\begin{equation}
 \mathfrak{C} =
  \bigl\{
    \xi_0 = \bar{\xi}_0 + i \, \breve{\xi}_0 \,\, \longrightarrow \,\, 
    \xi_\mathrm{end} = \bar{\xi}_\mathrm{end} + i \, \breve{\xi}_0  \bigr\},
\label{Pcontour}
\end{equation}
joining the points $\xi_0$ and $\xi_\mathrm{end}$ in $\mathbb{C}$. The constant imaginary part $\breve{\xi}_0$ of the complex distance $\xi \in \mathfrak{C}$ is usually taken to lie in the interval $\left[10^{-9},\, 10^{-3}\right]$. The real part $\bar{\xi}_\mathrm{end}$ of the complex terminal point $\xi_\mathrm{end}$ is taken here equal to $\bar{\xi}_\mathrm{end} \simeq 2 \, \bar{\Xi}_1$.
Thus the Lane--Emden function $\theta_{00}$ becomes complex-valued function in one complex variable, $\theta_{00}[\mathbb{I}_{\xi0} \subset \mathbb{C}] \subset \mathbb{C}$. Likewise, for the functions $\Theta_\sigma$ and $\Upsilon_\sigma$ (Eqs.~(\ref{T6501})--(\ref{T6502})) we write $\Theta_\sigma[\mathbb{I}_{\xi0} \subset \mathbb{C}] \subset \mathbb{C}$ and $\Upsilon_\sigma[\mathbb{I}_{\xi0} \subset \mathbb{C}] \subset \mathbb{C}$.
The initial conditions~(\ref{2.18}a,\,b) and (\ref{2.18TY}) become
\begin{equation}
\begin{aligned}
&\bar{\theta}_{00}(\xi_0) &=1, \qquad
&\bar{\theta}_{00}'(\xi_0) =0, \qquad 
&\breve{\theta}_{00}(\xi_0) = \breve{\theta}_{00}'(\xi_0) = 0, \\
&\bar{\Theta}_{\sigma}(\xi_0) &=1, \qquad 
&\breve{\Theta}_{\sigma}(\xi_0) =0, \qquad 
&\bar{\Upsilon}_{\sigma}(\xi_0) = \breve{\Upsilon}_{\sigma}(\xi_0) =0. 
\label{2.20e}
\end{aligned} 
\end{equation}
Furthermore, the initial conditions~(\ref{2.18}c,\,d) for the functions $\theta_{ij}$ with $i>0$ become
\begin{equation}
\theta_{ij}(\xi_0) = 
      \left( \bar{\theta}_{ij} \right)_0 + i \, 
      \left( \breve{\theta}_{ij} \right)_0.
\label{ivscomplex}
\end{equation} 
In detail, the real parts (\citep{GK14}, Eq.~(46)) are written as
\begin{equation}
\begin{aligned}
\left( \bar{\theta}_{ij} \right)_0 &=0, \qquad \, \,  
\left( \frac{d\bar{\theta}_{ij}}{d\xi} \right)_0 =0, \qquad i = 1, \,2, \, \, \, j = 0, \\ 
\left( \bar{\theta}_{ij} \right)_0 &=\xi^j, \qquad  
\left( \frac{d\bar{\theta}_{ij}}{d\xi} \right)_0 = j\,\xi^{j-1}, \qquad i = 1, \, 2, \, \, \, j \geq 2, \\
\left( \bar{\theta}_{30} \right)_0 &=0, \qquad \, \, 
\left( \frac{d\bar{\theta}_{30}}{d\xi} \right)_0 = 0, \\
\label{2.20e}
\end{aligned} 
\end{equation}
and the imaginary parts as (\citep{GK14}, Eq.~(47)) 
\begin{equation}
\left( \breve{\theta}_{ij} \right)_0 = 0, \qquad
\left( \frac{d\breve{\theta}_{ij}}{d\xi} \right)_0 = 0.
\label{2.20f}
\end{equation}

The raised-to-real-power terms involved in the definitions of the functions $S_{ij}$ (\citep{GK14}, Eq.~(13)) and in Eqs.~(\ref{T6501})--(\ref{T6502}) are written in terms of the auxiliary functions $\chi$ and $X = \Theta_\sigma^{1/N}$ as 
\begin{equation}
\begin{aligned}
&\theta_{00}^{n+1} = \chi^{N(n+1)}, \qquad \theta_{00}^n = \chi^{Nn}, \qquad \theta_{00}^{n-1} = \chi^{N(n-1)}, \qquad \theta_{00}^{n-2} = \chi^{N(n-2)},   \\
&\Theta_\sigma^{n+1} = X^{N(n+1)}, \quad \ \,
\Theta_\sigma^n = X^{Nn}, \quad \ \  
\Theta_\sigma' = NX^{N-1}X'.
\label{rtrpt} 
\end{aligned}
\end{equation}
For $n = 1.0$ and $2.0$, we choose $N = 1$; thus $\theta_{00} = \chi$ and $\Theta_\sigma = X$. For $n = 1.5$ and $2.5$, we take $N = 2$, which yields $\theta_{00}^{1.5} = \chi^3$, $\theta_{00}^{2.5} = \chi^5$, $\Theta_\sigma^{1.5} = X^3$, and $\Theta_\sigma^{2.5} = X^5$. Finally, for $n = 2.9$, we choose $N = 10$, which gives $\theta_{00}^{2.9} = \chi^{29}$ and $\Theta_\sigma^{2.9} = X^{29}$.

\section{Units}
In this study, the abbreviations ``cgs'', ``gu'', `pu'', and ``cpu'' denote ``cgs units'', ``gravitational units'', ``polytropic units related to the gravitational units'', and ``classical polytropic units'', respectively (for a discussion on the gravitational units and their related polytropic units, see e.g. \citep{GS11}, Sec.~1.2). The units of several physical quantities in the system of gravitational units are given in Table \ref{tab:ugu} and play the role of ``conversion coefficients'', which convert a physical measure in gu to the respective measure in cgs. For instance, if the measure of a density in gu is $\rho_\mathrm{gu}$, then $\rho_\mathrm{cgs} = [D]_\mathrm{gu} \, \rho_\mathrm{gu}$ is its measure in cgs. In gu, any physical quantity has a dimension of the form $\mathrm{cm}^\gamma$ (\citep{GS11}, Eq.~(1); there is only one base unit in gu, the length, measured in cm), that is, explicitly, it has a dimension $[\gamma]$. If $\gamma=0$ for a particular physical quantity, then this quantity is dimensionless in gu. 

The units of several physical quantities in the system of the polytropic units related to the gravitational units (see e.g. \citep{CST94}, Eqs.~(4)--(13)) are given in Table \ref{tab:upu} and convert a physical measure in pu to the respective measure in gu. For example, if the measure of a density in pu is $\rho_\mathrm{pu}$, then $\rho_\mathrm{gu} = [D]_\mathrm{pu} \, \rho_\mathrm{pu}$ is its measure in gu; accordingly, its measure in cgs is $\rho_\mathrm{cgs} = [D]_\mathrm{gu} [D]_\mathrm{pu} \, \rho_\mathrm{pu}$. All physical quantities are dimensionless in pu, since their physical dimensions are assigned to their respective units.  

The polytropic units related to the gravitational units should not be confused with the classical polytropic units (see e.g. \citep{GTV79}, Sec.~8), defined on the basis of the normalization equations~(\ref{2.5bc}a,\,b). The units of several physical quantities in the system of classical polytropic units are given in Table \ref{tab:cpu} and play the role of conversion coefficients, which convert a physical measure in cpu to the respective measure in cgs. For example, if the measure of a density in cpu is $\rho_\mathrm{cpu}$, then $\rho_\mathrm{cgs} = [D]_\mathrm{cpu} \, \rho_\mathrm{cpu}$ is its measure in cgs; accordingly, its measure in gu is $\rho_\mathrm{gu} = (1/[D]_\mathrm{gu}) \, [D]_\mathrm{cpu} \, \rho_\mathrm{cpu}$ and its measure in pu is $\rho_\mathrm{pu} = (1/[D]_\mathrm{pu}) \, (1/[D]_\mathrm{gu}) \, [D]_\mathrm{cpu} \, \rho_\mathrm{cpu}$. All physical quantities are dimensionless in cpu, since their physical dimensions are assigned to their respective units.

In almost all the computations of this study, we use cpu measures of physical quantities and characteristics, since PNA is inherently oriented to cpu. However, since pu is the system mostly used in the bibliography, all results and comparisons are quoted in pu.

\begin{table}
\begin{center}
\caption{Units of several physical quantities in the system of ``gravitational units'' (gu),  converting physical measures in gu to respective measures in cgs.\label{tab:ugu}}
\begin{tabular}{rllr} 
\hline\hline
physical quantity  & dimension of the     & value of          & numeric value         \\
and its unit in gu & quantity in gu       & the unit          & of the unit           \\
\hline
Length, $[L]_\mathrm{gu}$                  & $\mathrm{cm}^1$      & $1$                & $1.000(+00)$ \\
Mass, $[M]_\mathrm{gu}$                    & $\mathrm{cm}^1$      & $c^2/G$            & $1.347(+28)$ \\
Density, $[D]_\mathrm{gu}$                 & $\mathrm{cm}^{-2}$   & $c^2/G$            & $1.347(+28)$ \\
Pressure, $[P]_\mathrm{gu}$                & $\mathrm{cm}^{-2}$   & $c^4/G$            & $1.210(+49)$ \\
Energy, $[T]_\mathrm{gu}=[W]_\mathrm{gu}$  & $\mathrm{cm}^1$      & $c^4/G$            & $1.210(+49)$ \\
Angular velocity, $[\Omega]_\mathrm{gu}$   & $\mathrm{cm}^{-1}$ & $c$                  & $2.998(+10)$ \\
Angular momentum, $[J]_\mathrm{gu}$        & $\mathrm{cm}^{2}$  & $c^3/G$            & $4.038(+38)$ \\
Moment of inertia, $[I]_\mathrm{gu}$       & $\mathrm{cm}^{3}$  & $c^2/G$            & $1.347(+28)$ \\
\hline
\end{tabular}
\end{center}
\end{table}

\begin{table}
\begin{center}
\caption{Units of several physical quantities in the system of ``polytropic units related to the gravitational units'' (pu), used to convert physical measures in pu to respective measures in gu. The symbol $K$ denotes here the measure of the polytropic constant in gu, $K = K_\mathrm{gu}$.\label{tab:upu}}
\begin{tabular}{rl} 
\hline\hline
physical quantity  & value of          \\
and its unit in pu & the unit          \\
\hline
Length, $[L]_\mathrm{pu}$                  & $K^{n/2}$     \\
Mass, $[M]_\mathrm{pu}$                    & $K^{n/2}$     \\
Density, $[D]_\mathrm{pu}$                 & $K^{-n}$      \\
Pressure, $[P]_\mathrm{pu}$                & $K^{-n}$      \\
Energy, $[T]_\mathrm{pu}=[W]_\mathrm{pu}$  & $K^{-n}$      \\
Angular velocity, $[\Omega]_\mathrm{pu}$   & $K^{-n/2}$    \\
Angular momentum, $[J]_\mathrm{pu}$        & $K^n$         \\
Moment of inertia, $[I]_\mathrm{pu}$       & $K^{3n/2}$    \\
\hline
\end{tabular}
\end{center}
\end{table}

\begin{table}
\begin{center}
\caption{Units of several physical quantities in the system of ``classical polytropic units'' (cpu), used to convert physical measures in cpu to respective measures in cgs.
\label{tab:cpu}}
\begin{tabular}{rl} 
\hline\hline
physical quantity   & value of          \\
and its unit in cpu & the unit          \\
\hline
Length, $[L]_\mathrm{cpu}$                  & $\alpha$~~(see Eq.~(\ref{2.5bc}b))              \\
Density, $[D]_\mathrm{cpu}$                 & $\rho_\mathrm{c}$~(see Eq.~(\ref{2.5bc}a))      \\
Pressure, $[P]_\mathrm{cpu}$                & $K_\mathrm{cgs} \, \rho_\mathrm{c}^\Gamma$~(see Eq.~(\ref{2.5}))      \\
Mass, $[M]_\mathrm{cpu}$                    & $4 \, \pi \, \alpha^3 \, \rho_\mathrm{c}$ \\
Energy, $[T]_\mathrm{cpu}=[W]_\mathrm{cpu}$ & $16 \, \pi^2 \, G \, \alpha^5 \, \rho_\mathrm{c}^2$      \\
Angular velocity, $[\Omega]_\mathrm{cpu}$   & $4 \, \pi \, G \, \rho_\mathrm{c}$    \\
Angular momentum, $[J]_\mathrm{cpu}$        & $8 \, \pi^{1.5} \, G^{0.5} \, \alpha^5 \, \rho_\mathrm{c}^{1.5}$         \\
Moment of inertia, $[I]_\mathrm{cpu}$       & $4 \, \pi \, \alpha^5 \, \rho_\mathrm{c}$    \\
\hline
\end{tabular}
\end{center}
\end{table}

\section{The Computations, I}
\label{comp}
Preliminaries regarding the computational environment used in this work and the Fortran code \texttt{DCRKF54} \citep{GV12} for solving complex IVPs can be found in \citep{GK14}.

Our code runs in four steps. Step 1 (S1) solves Eqs.~(\ref{2.10}) for the functions $\theta_{00}, \, \theta_{10}, \, \theta_{12}, \, \theta_{14}, \, \Theta_\sigma, \, \Upsilon_\sigma$, and stores the solution into proper arrays. All these arrays are interpolated by cubic splines in both their real and imaginary parts. All interpolations have as independent variable the real part $\bar{\xi}$ of the complex distance $\xi$. The radius $\bar{\Xi}_1$ of the undistorted configuration is computed as the first root of the algebraic equation
\begin{equation}
F_\mathrm{I} \left[ \bar{\theta}_{00} \right] \left( \bar{\xi} \, \right) = 0,
\end{equation}
where $F_\mathrm{I} \left[ \bar{\theta}_{00} \right]$ is the interpolating function for the real part $\bar{\theta}_{00}$ of the function $\theta_{00}$. 
Then S1 calculates the surface values of the real and imaginary parts of all functions computed and the parameters $k_{00}$, $k_{10}$, $c_{00}$, $c_{10}$, $A_{12}$.
Next, Step 2 (S2) solves Eqs.~(\ref{2.10}) for the functions $\theta_{20}$, $\theta_{22}$, $\theta_{24}$, storing their values into proper arrays. All arrays are interpolated by cubic splines in both their real and imaginary parts. 

Step 3 (S3) proceeds with a scheme able to compute the function $\Theta(\bar{\xi},\,\mu)$ 
at any point $(\bar{\xi},\,\mu)$, with $\bar{\xi} \leq 2 \bar{\Xi}_1$, lying either inside or outside the nonrotating Newtonian configuration of radius $\bar{\Xi}_1$. On the basis of this scheme, S3 can compute the surface of the configuration, that is, the root $\bar{\Xi}_\mu$ of the equation 
\begin{equation}
F_\mathrm{I}\left[\bar{\Theta}\right](\bar{\xi},\,\mu) =
\sum_{i=0,\,2}^4 F_\mathrm{I}\left[\bar{\Theta}_i\right](\bar{\xi}\,) \, P_i(\mu) = 0,
\label{boundR}
\end{equation}
where $F_\mathrm{I} \left[ \bar{\Theta}_i \right]$ are the interpolating functions for the real parts $\bar{\Theta}_i$ of the functions $\Theta_i$ defined by Eq.~(\ref{2.9a}), at any $\mu$ with a given accuracy $\tau$.

In the framework of HAS, the boundary of the configuration is assumed to coincide with the equidensity surface
\begin{equation}
\left| F_\mathrm{I}[\bar{\Theta}](\bar{\Xi}_\mu,\,\mu)  \right| = 
                    \tau_\mathrm{s} > 0,
\label{equidensity} 
\end{equation}
where  $\tau_\mathrm{s}$ is a given ``surface parameter'' (for a similar issue arising in the framework of the well-known Hartle's perturbation method, see \citep{GK08}, Sec.~5.1). In addition, it is assumed that the function $F_\mathrm{I}[\bar{\Theta}]$ approaches the boundary condition~(\ref{equidensity}) from positive values, $F_\mathrm{I}[\bar{\Theta}] > 0$, in the case of the highly stiff EOS $n=1.0$,
\begin{equation} 
0 \leq F_\mathrm{I}[\bar{\Theta}](\bar{\Xi}_\mu,\,\mu) - \tau_\mathrm{s} < \tau,
\label{Fgtt}
\end{equation} 
and from negative values, $F_\mathrm{I}[\bar{\Theta}] < 0$, in the case of the moderately stiff and soft EOSs with $n \geq 1.5$,
\begin{equation}
0 \leq -F_\mathrm{I}[\bar{\Theta}](\bar{\Xi}_\mu,\,\mu) - \tau_\mathrm{s} < \tau.
\label{Fltt}
\end{equation}
It is worth mentioning here that, among the members of a collection of EOSs, the EOS deriving the larger $p$ for a given $\rho$ is the stiffest EOS in the collection; while the EOS leading to the smaller $p$ for the same $\rho$ is the softest EOS. For increasing $n$, the polytropic EOSs are getting softer; thus, in the collection $n = 1.0, \, 1.5, \, 2.0, \, 2.5, \, 2.9$, stiffest EOS is that with $n = 1.0$, while softest EOS is that with $n = 2.9$.  

To compute the critical rotation parameter $\upsilon_\mathrm{c}$, S3 treats Eq.~(\ref{boundR}) in the full form of its dependencies,
\begin{equation}
\left|
\sum_{i=0,\,2}^4 F_\mathrm{I}\left[\bar{\Theta}_i\right](\sigma, \, \upsilon, \,
                                   \bar{\Xi}_\mathrm{e}\,) \, P_i(\mu = 0) 
\right| = \tau_\mathrm{s}, 
\label{boundRSU} 
\end{equation}  
and solves this equation for the ``root'' $\upsilon$ when the ``variables'' $\sigma$ and $\bar{\Xi}_\mathrm{e}$ are given. Accordingly, the root $\upsilon(\sigma, \, \bar{\Xi}_\mathrm{e})$ is the rotation parameter for which the distorted configuration obtains equatorial radius $\bar{\Xi}_\mathrm{e}$ under gravitation parameter $\sigma$.  Solving Eq.~(\ref{boundRSU}) with a given accuracy $\tau$ for a mesh of values $\left\{ \left( \bar{\Xi}_\mathrm{e} \right)_m \right\}$ lying in an appropriate interval $\mathbb{I}(\bar{\Xi}_\mathrm{e})$ --- say $\mathbb{I}\left(\bar{\Xi}_\mathrm{e}\right) = [1.2 \, \bar{\Xi}_1, 1.8 \, \bar{\Xi}_1]$ --- and constructing the interpolating function $F_\mathrm{I}\left[\upsilon\right]\left(\bar{\Xi}_\mathrm{e}\right)$, S3 localizes the maximum value $\upsilon_\mathrm{c}$ of this function. This maximum represents the respective $\upsilon_\mathrm{c}$ for the particular $\sigma$; and the value $\bar{\Xi}_\mathrm{e}$ deriving $\upsilon_\mathrm{c}$ is the equatorial radius under gravitation parameter $\sigma$ and rotation parameter $\upsilon_\mathrm{c}$.

By studying the variation of $\upsilon_\mathrm{c}$ with the surface parameter $\tau_\mathrm{s}$, the latter written as $\tau_\mathrm{s}=\sigma/\nu$ with $\nu=1,\,2,\,\dots$, we can determine an optimum value for $\tau_\mathrm{s}$. In particular, our numerical experiments show that there is a value $\nu$, about which this variation changes from near quadratic to near linear of small slope. Such a change occurs when $\nu \sim 10$; hence, the value $\tau_\mathrm{s} \sim \sigma/10$ is adopted in the present study as an optimum $\tau_\mathrm{s}$. Note that in \citep{GK14} all models resolved have gravitation parameters $\sigma \leq 0.008$; due to such small values of $\sigma$, the surface parameter $\tau_\mathrm{s}$ is taken to be zero in \citep{GK14}.

\section{Physical Characteristics}
\label{physchar}
Since physical interest focuses on real parts of functions and parameters, we will hereafter quote only such values and, for simplicity, we will drop overbars denoting real parts of complex quantities. Second, for brevity, we will denote the interpolating functions by the symbols denoting so far the respective mathematical functions; for example, we will write $\Theta$ in the place of the respective interpolating function $F_\mathrm{I}\left[\Theta\right]$ defined by Eq.~(\ref{boundR}). Third, any symbol not explicitly connected to a system of units will denote the cgs measure of the respective physical characteristic; for example, the symbol $M$ will denote the cgs measure of the gravitational mass.    
  
In the framework of PNA, the Newtonian relations for the physical characteristics of interest are modified as follows. First, the gravitational mass $M$ is given by (cf. \citep{GTV79}, Eq.~(8.2))
\begin{equation}
M = \int_V E \, dV = 
[M]_\mathrm{cpu}
\int^{1}_{0} \int^{\xi_\mathrm{t}}_{0} \Psi^n \, \xi^2 \, d\xi \, d\mu,  
\label{eqmass}
\end{equation}
where $dV$ is the coordinate volume element, $\xi_\mathrm{t}$ the upper limit of the integration in $\xi$ chosen so that $\Xi_\mathrm{e} < \xi_\mathrm{t} \leq \xi_\mathrm{end}$, $E$ the mass-energy density, and $\Psi^n$ the cpu measure of $E$, i.e. $E_\mathrm{cpu} = \Psi^n$. The functions $\Theta$ and $\Psi$ are connected via a sequence of equations, which are based on the relation (cf.~\citep{GS11}, Eq.~(6))
\begin{equation}
E_\mathrm{gu} = \rho_\mathrm{gu} + n \, p_\mathrm{gu}
\label{Erho}
\end{equation}
holding in gu. To find $\Psi(\Theta)$, we first calculate $E_\mathrm{gu}$,
\begin{equation}
\rho = \rho_\mathrm{c} \, \Theta^n,           \ \ 
\rho_\mathrm{gu} = \rho / [D]_\mathrm{gu},    \ \
p_\mathrm{gu} = K_\mathrm{gu} \, \rho_\mathrm{gu}^\Gamma,  \ \
E_\mathrm{gu} = \rho_\mathrm{gu} + n \, p_\mathrm{gu},
\label{firstseq} 
\end{equation}
where $K_\mathrm{gu}$ is the measure in gu of the polytropic constant $K$ (\citep{GS11}, Sec.~1.2),
\begin{equation}
K_\mathrm{gu} = \left([M]_\mathrm{gu}^\Gamma / [P]_\mathrm{gu}\right) K.
\label{Kappa}
\end{equation}  
Next, we convert measures back to cpu,
\begin{equation}
E = [D]_\mathrm{gu} \, E_\mathrm{gu}, \ \
E_\mathrm{cpu} = E / \rho_\mathrm{c}, \ \
\Psi^n = E_\mathrm{cpu},                           \ \
\Psi = \left(E_\mathrm{cpu}\right)^{1/n}.
\label{secondseq}
\end{equation} 
 
The baryonic mass $M_0$, also called rest mass, is given by (cf.~\citep{GS11}, Eq.~(108))  
\begin{equation}
M_0 = \int_\mathcal{V} \rho \, d\mathcal{V} = 
[M]_\mathrm{cpu} \,
\int^{1}_{0} \int^{\xi_\mathrm{t}}_{0} \Theta^n \, \xi^2 \, \Lambda \, d\xi \, d\mu,  
\label{eqMass0}
\end{equation}
In this relation, we use the ansatz
\begin{equation}
dV \rightarrow d\mathcal{V},
\label{dVansatz}
\end{equation}
with the meaning that the coordinate volume element $dV$ is substituted by the proper volume element $d\mathcal{V}$ (a discussion on this matter for nonrotating relativistic objects can be  found in \citep{B11}, Sec.~2). The ansatz~(\ref{dVansatz}) is equivalent to the substitution
\begin{equation}
d\xi \rightarrow \Lambda \, d\xi 
\label{dxiansatz}
\end{equation} 
of the coordinate differential $d\xi$, where the function $\Lambda$ plays the role of the metric function $e^{\lambda/2}$ (see e.g. \citep{GK08}, Eqs.~(1) and (5)) in the case that a configuration suffers both relativistic and rotational distortions,
\begin{equation}
\Lambda(\sigma, \, \upsilon, \, \xi) = \left[ 1 - \frac{2 \, G \, 
           m(\sigma, \, \upsilon, \, \xi)}
       {c^2 \, \alpha \, \xi_\mathrm{avr}(\xi)} \right]^{-1/2};
\label{Lambda}
\end{equation} 
the term $4 \, \pi \, \alpha^2$, which is also involved in the expression for $\Lambda$ (see e.g. \citep{B11}, Sec.~2), is incorporated into the respective cpu units. The meaning of $\Lambda(\sigma, \, \upsilon, \, \xi)$ is that, for a HAS solution of given $\sigma$ and $\upsilon$, the metric function $\Lambda$ can be considered as a function of the coordinate $\xi$, playing here the role of the semimajor axis $\xi_\mathrm{e}$ of a spheroidal equidensity surface of density $\Psi^n(\xi_\mathrm{e} = \xi,\,\mu=0)$. Accordingly, the function $m(\sigma, \, \upsilon, \, \xi)$ is the gravitational mass inside this spheroid, given by 
\begin{equation}
m(\sigma, \, \upsilon, \, \xi) =  
[M]_\mathrm{cpu}
\int^{1}_{0} \int^{\xi_\mathrm{t}, \, \Psi^n \geq \Psi^n(\xi_\mathrm{e}=\xi, \, \mu=0)}_{0} 
             \Psi^n \, \xi^2 \, d\xi \, d\mu,  
\label{eqpartmass}
\end{equation} 
where in this integration participate only the mass elements with densities $\Psi^n \geq \Psi^n(\xi_\mathrm{e}=\xi, \, \mu=0)$. The function $\xi_\mathrm{avr}(\xi)$ in Eq.~(\ref{Lambda}) denotes the average radius of the particular spheroid; note that, if $\xi_\mathrm{p}$ is its semiminor axis,    
\begin{equation}
\Psi^n(\xi_\mathrm{p}, \, \mu=1) = \Psi^n(\xi_\mathrm{e}=\xi, \, \mu=0)
\end{equation}
then a rough approximation of $\xi_\mathrm{avr}(\xi)$ is 
\begin{equation}
\xi_\mathrm{avr}(\xi) \, \approx \, (\xi_\mathrm{e} + \xi_\mathrm{p})/2.
\end{equation}

Next, the proper mass $M_\mathrm{P}$ (cf.~\citep{GS11}, Eq.~(109)) is written as
\begin{equation}
M_\mathrm{P} = \int_\mathcal{V} E \, d\mathcal{V} = 
[M]_\mathrm{cpu}
\int^{1}_{0} \int^{\xi_t}_{0} \Psi^n \, \xi^2 \, \Lambda \, d\xi \, d\mu,  
\label{eqMassP}
\end{equation}
It is worth remarking here that the only difference between Eqs.~(\ref{eqMass0}) and (\ref{eqMassP}) is the appearance of the mass-energy density $\Psi^n$ in the place of the rest-energy density $\Theta^n$.

The rotational kinetic energy $T$ is given by (cf.~\citep{GTV79}, Eq.~(8.5))
\begin{equation}
T = 
\int_\mathcal{V} E \, \mathbf{v} \cdot \mathbf{v} \, d\mathcal{V} =  
[T]_\mathrm{cpu} \, \left[ \, \frac{1}{2} \, \, \Omega_{*}^2 \, \, 
\int^{1}_{0} \int^{\xi_\mathrm{t}}_{0} (1 - \mu^2) \, \Psi^n \, 
                              \xi^4 \, \Lambda \, d\xi \, d\mu \right],  
\label{eqkin}
\end{equation}
where $\Omega_*$ is the measure in cpu of the angular velocity $\Omega$; 
thus, in cgs, $\Omega = [\Omega]_\mathrm{cpu} \, \Omega_*$. Combining Eq.~(\ref{2.7}a) with the definition of $[\Omega]_\mathrm{cpu}$ (Table \ref{tab:cpu}, sixth entry), we can verify that (\citep{GTV79}, Eq.~(2.20))
\begin{equation}
\Omega_* = \sqrt{\frac{\upsilon}{2}}.
\label{Omega*}
\end{equation}

The gravitational potential energy $W$ is written as (cf.~\citep{GTV79}, Eq.~(8.6))
\begin{equation}
W = 
\int_\mathcal{V} E \, \Phi \, d\mathcal{V} =  
[W]_\mathrm{cpu} \, \int^{1}_{0} \int^{\xi_\mathrm{t}}_{0} 
\Psi^n \, \Phi \, \xi^2 \, \Lambda \, d\xi \, d\mu,  
\label{eqpot}
\end{equation}
where $[W]_\mathrm{cpu} = [T]_\mathrm{cpu}$ (Table \ref{tab:cpu}, fifth entry), and the involved gravitational potential $\Phi$ is defined by (cf.~\citep{H86}, Eq.~(2)) 
\begin{equation}
\Phi(\mathbf{r}) = \, - \, G \,
\int_{\mathcal{V}'} \frac{E(\mathbf{r'})}
     {\left| \mathbf{r} - \mathbf{r}' \right|} \, \, d\mathcal{V}'. 
\label{eqgp}
\end{equation}

The angular momentum $J$ is given by (cf.~\citep{GTV79}, Eq.~(8.7))
\begin{equation}
J = 
\int_\mathcal{V} E \, \mathbf{r} \times \mathbf{v} \, d\mathcal{V} = 
[J]_\mathrm{cpu} \, \left[ \Omega_{*} \, \int^{1}_{0} \int^{\xi_\mathrm{t}}_{0} 
           (1 - \mu^2) \, \Psi^n \, \xi^4 \, \Lambda \, d\xi \, d\mu \right].  
\label{eqagmom}
\end{equation}

Finally, it is worth mentioning that the moment of inertia $I$ is given by (see e.g. \citep{GK08}, Eq.~(22))
\begin{equation}
I = \frac{J}{\Omega} = \, [I]_\mathrm{cpu} \, \left( \frac{J_\mathrm{cpu}}
                                                          {\Omega_*} 
                                         \right).
\label{eqmoi}
\end{equation}

\section{The Computations, II}
\label{comp2}
The physical characteristics discussed in Sec.~\ref{physchar} are computed by passing certain quantities found by S3 to the next Step 4 (S4). This step integrates all double integrals involved in the definitions of the physical characteristics by using Simson's formula as proposed and described by Hachisu (\citep{H86}, Sec.~IV).
 
In detail, we first define two coordinate arrays; namely, the array $\{\mu_i\}$ in the $\mu$-direction (cf.~\citep{H86}, Eq.~(51)),
\begin{equation}
\mu_i = (i-1)/(\mathtt{KAP} -1 ), \qquad i = 1, \, 2, \, \dots, \, \mathtt{KAP};
\label{KAP} 
\end{equation}
and the array $\{\xi_j\}$ in the $\xi$-direction (cf.~\citep{H86}, Eq.~(50)),
\begin{equation}
\xi_j = \left[ (j-1)/(\mathtt{KRP} -1) \right]\,\xi_\mathrm{t}, \qquad 
j = 1, \, 2, \, \dots, \, \mathtt{KRP}.
\label{KRP} 
\end{equation}
As explained in Sec.~\ref{physchar}, $\xi_\mathrm{t}$ is the upper limit of the integrations with respect to the coordinate $\xi$, lying in the interval $\mathbb{I}(\xi_\mathrm{t}) = (\Xi_\mathrm{e}, \, \xi_\mathrm{end}]$. In the present study, the ``number of the elements $\mu_i$'' \texttt{KAP} and the ``number of the elements $\xi_j$'' \texttt{KRP} are taken equal to $\mathtt{KAP} = \mathtt{KRP} = 201$; and the upper limit of the integrations in the coordinate $\xi$ is taken equal to $\xi_\mathrm{t} = 1.125 \, \, \Xi_\mathrm{e}$.

Having defined the coordinate arrays, we proceed with the computation of the array $\{\Theta^n_{i,j}\}$, which has as elements the rest-mass densities $\Theta^n_{i,j}=\Theta(\xi_j,\,\mu_i)^n$, 
\begin{equation}
\Theta^n_{i,j} = \Theta(\xi_j,\,\mu_i)^n \ \ \ \mathrm{if} \ \ \Theta(\xi_j,\,\mu_i) > 0; 
\ \mathrm{else} \  \ \Theta^n_{i,j} = 0.
\label{arrayTheta} 
\end{equation}
Likewise, the array $\{\Psi^n_{i,j}\}$ with elements the mass-energy densities $\Psi^n_{i,j}=\Psi(\xi_j,\,\mu_i)^n$ is given by
\begin{equation}
\Psi^n_{i,j} = \Psi(\xi_j,\,\mu_i)^n \ \ \ \mathrm{if} \ \ \Psi(\xi_j,\,\mu_i) > 0; 
\ \mathrm{else} \  \ \Psi^n_{i,j} = 0.
\label{arrayPsi} 
\end{equation}
To calculate the values $\Psi(\xi_j,\,\mu_i)$ from the values $\Theta(\xi_j,\,\mu_i)$, we use the relations~(\ref{firstseq})--(\ref{secondseq}).

Now, to compute the gravitational mass $M$, we first construct an auxiliary array $\{Q_j\}$ with elements
\begin{equation}
Q_j = \, \sum_{i=1(2)}^\mathtt{KAP-2} \, \frac{1}{6} \, 
\left( \mu_{i + 2} - \mu_i \right) \,  
\left[ \Psi^n_{i,j}  + 4 \, \Psi^n_{i + 1,j} + \Psi^n_{i + 2, j} 
\right].
\end{equation}
Then $M$ results from the relation 
\begin{equation}
M = [M]_\mathrm{cpu} \,  
\sum_{j=1(2)}^\mathtt{KRP-2} \, \frac{1}{6} \, 
(\xi_{i + 2} - \xi_{i}) \, \left[
\xi_{j}^2 \, Q_{j} + 4 \, \xi_{j + 1}^2 \, Q_{j + 1} + \xi_{j+2}^2 \, Q_{j+2} \right].
\label{gravmass} 
\end{equation}  

Next, to compute the array $\{m_k\}$ with elements $m_k = m(\sigma,\,\upsilon,\,\xi_k)$ (Eq.~(\ref{eqpartmass})), we first construct an auxiliary array $\{q_{j,k}\}$ with elements
\begin{equation}
q_{j,k} = \, \sum_{i=1(2)}^\mathtt{KAP-2} \, \frac{1}{6} \, 
\Delta_{i+2,j,k} \left( \mu_{i + 2} - \mu_i \right) \,  
\left[ \Psi^n_{i,j}  + 4 \, \Psi^n_{i + 1,j} + \Psi^n_{i + 2, j} 
\right].
\end{equation}
where
\begin{equation}
\Delta_{i,j,k} = 1 \ \ \ \mathrm{if} \ \ \Psi^n_{i,j} \geq \Psi^n_{1,k};
\ \mathrm{else} \ \ \Delta_{i,j,k} = 0.
\end{equation}
Then the elements $m_k$ are computed by 
\begin{equation}
m_k = [M]_\mathrm{cpu} \, 
\sum_{j=1(2)}^\mathtt{KRP-2} \, \frac{1}{6} \,  
(\xi_{i + 2} - \xi_{i}) \, \left[
\xi_{j}^2 \, q_{j,k} + 4 \, \xi_{j + 1}^2 \, q_{j + 1,k} + \xi_{j+2}^2 \, q_{j+2,k} 
                           \right].
\label{partialmasses} 
\end{equation}

To proceed with the computation of the array $\{\Lambda_k\}$ with elements $\Lambda_k = \Lambda(\sigma,\,\upsilon,\, \xi_k)$ (Eq.~(\ref{Lambda})), we first construct the array $\{\xi_{\mathrm{(avr)}k}\}$ with elements $\xi_{\mathrm{(avr)}k} = \xi_\mathrm{avr}(\xi_k)$. Using the names \texttt{XI(K)} for $\xi_k$, \texttt{XI\_AVR(K)} for $\xi_{\mathrm{(avr)}k}$, \texttt{PSI\_N(I,J)} for $\Psi^n_{i,j}$, we compute the element(s) $\xi_{\mathrm{(avr)}k}$ by the code
\begin{footnotesize}
\begin{verbatim}
      XI_AVR(K)=XI(K)
      LOOP_I: DO I=2,KAP
          LOOP_J: DO J=1,KRP
              IF (PSI_N(I,J) < PSI_N(1,K)) THEN
                  XI_SURFACE_I=XI(J-1)
                  EXIT LOOP_J
              END IF
          END DO LOOP_J
          XI_AVR(K)=XI_AVR(K)+XI_SURFACE_I
      END DO LOOP_I
      XI_AVR(K)=XI_AVR(K)/KAP
\end{verbatim}  
\end{footnotesize}   
Then the elements $\Lambda_k$ are computed by (Eq.~(\ref{Lambda}))
\begin{equation}
\Lambda_k = \left[ 1 - \frac{2 \, m_k / [M]_\mathrm{gu}}{\alpha \, \xi_{\mathrm{(avr)}k}}
            \right]^{-1/2}.
\end{equation}

Next, to compute the baryonic mass $M_0$, we reconstruct the auxiliary array $\{Q_j\}$ with new elements
\begin{equation}
Q_j = \, \Lambda_j \, \sum_{i=1(2)}^\mathtt{KAP-2} \, \frac{1}{6} \,  
\left( \mu_{i + 2} - \mu_i \right) \,  
\left[ \Theta^n_{i,j}  + 4 \, \Theta^n_{i + 1,j} + \Theta^n_{i + 2, j} 
\right].
\label{QM0}
\end{equation}
Then $M_0$ is computed by the relation~(\ref{gravmass}) with the elements $Q_j$ of Eq.~(\ref{QM0}).
Likewise, to compute the proper mass $M_\mathrm{P}$, we reconstruct the auxiliary array $\{Q_j\}$ with new elements
\begin{equation}
Q_j = \, \Lambda_j \, \sum_{i=1(2)}^\mathtt{KAP-2} \, \frac{1}{6} \, 
\left( \mu_{i + 2} - \mu_i \right) \,  
\left[ \Psi^n_{i,j}  + 4 \, \Psi^n_{i + 1,j} + \Psi^n_{i + 2, j} 
\right],
\label{QMP}
\end{equation}
and we compute $M_\mathrm{P}$ by the relation~(\ref{gravmass}) with the  elements $Q_j$ of Eq.~(\ref{QMP}).

We proceed now with the computation of the rotational kinetic energy $T$. First, we reconstruct the auxiliary array $\{Q_j\}$ with new elements  
\begin{equation}
\begin{aligned}
Q_j = & \, \, \Lambda_j \,  
\sum_{i=1(2)}^\mathrm{KAP-2} \, \frac{1}{6} \, (\mu_{i + 2} - \mu_{i}) \\ 
      & \times \, \left[ 
\Psi_{i,j} \, (1 - \mu_{i}^2) + 4 \, \Psi_{i + 1, j} \, (1-\mu_{i + 1}^2) + 
\Psi_{i + 2, j} \, (1 - \mu_{i + 2}^2) 
                  \right].
\end{aligned}
\label{QforTJ}
\end{equation}
Next, we compute $T$ by the relation
\begin{equation}
T= \, [T]_\mathrm{cpu} \, \frac{1}{2} \, \Omega_*^2 \, 
\sum_{j=1(2)}^\mathrm{KRP-2} \, \frac{1}{6} \, 
(\xi_{i + 2} - \xi_{i}) \left[
\xi_{j}^4 \, Q_j + 4 \, \xi_{j + 1}^4 \, Q_{j + 1} + \xi_{j+2}^4 \, Q_{j+2}
                        \right]
\label{Trot} 
\end{equation} 
by using the elements $Q_j$ of Eq.~(\ref{QforTJ}).

In order to compute the gravitational potential energy $W$, we need first to construct the auxiliary arrays $\{Q_{k,\ell}\}$ with elements (cf.~\citep{H86}, Eq.~(54))
\begin{equation}
\begin{aligned}
Q_{k,\ell} = & \, \, \Lambda_k \, 
\sum_{i=1(2)}^\mathrm{KAP-2} \, \frac{1}{6} \, (\mu_{i + 2} - \mu_{i}) \\ 
      & \times \, \left[ 
\Psi_{i,k} \, P_{2\ell}(\mu_{i}) + 4 \, \Psi_{i + 1, k} \, P_{2\ell}(\mu_{i + 1}) + 
\Psi_{i + 2, k} \, P_{2\ell}(\mu_{i + 2}) 
                  \right],
\end{aligned}
\end{equation}
and $\{R_{\ell,j}\}$ with elements (cf.~\citep{H86}, Eq.~(55))
\begin{equation}
\begin{aligned}
R_{\ell,j} = & \sum_{k=1(2)}^\mathrm{KRP-2} \, \frac{1}{6} \, (\xi_{k + 2} - \xi_{k}) \\ 
             & \times \, \left[ 
Q_{k,\ell} \, f_{2\ell}(\xi_{k},\xi_j) + 
4 \, Q_{k + 1,\ell} \, f_{2\ell}(\xi_{k + 1},\xi_j) + 
Q_{k + 2,\ell} \, f_{2\ell}(\xi_{k + 2},\xi_j) 
                         \right],
\end{aligned}
\end{equation}
where the functions $f_{2\ell}(\xi_j,\xi_k)$ are defined by Eq.~(3) of \citep{H86}. Then the elements $\Phi_{i,j}$ of the array $\{\Phi_{i,j}\}$ are given by (cf.~\citep{H86}, Eqs.~(2) and (56); the coefficient $4 \pi G$ has been incorporated into the respective units)
\begin{equation}
\Phi_{i,j} = - \sum_{\ell=0}^{\mathtt{KPL}} R_{\ell,j} \, P_{2\ell}(\mu_{i}).
\end{equation}
In the present study, the ``cutoff number of the Legendre polynomials'' \texttt{KPL} is taken equal to $\mathtt{KPL} = 8$; so, we use Legendre polynomials up to $P_{16}(\mu)$. It remains to construct the auxiliary array $\{S_j\}$ with elements (cf.~\citep{H86}, Eq.~(59))
\begin{equation}
S_j = \, \Lambda_j \, \sum_{i=1(2)}^\mathtt{KAP-2} \, \frac{1}{6} \, 
\left( \mu_{i + 2} - \mu_i \right) \,  
\left[ \Psi^n_{i,j} \, \Phi^n_{i,j} + 
4 \, \Psi^n_{i + 1,j} \, \Phi^n_{i + 1,j} + \Psi^n_{i + 2, j} \, \Phi^n_{i + 2, j} 
\right],
\label{SMP}
\end{equation}
Then $|W|$ results from the relation (cf.~\citep{H86}, Eq.~(60); the coefficient $2\pi$ has been incorporated into the respective units)
\begin{equation}
|W| = [W]_\mathrm{cpu} \, \left| \,   
-\sum_{j=1(2)}^\mathtt{KRP-2} \, \frac{1}{6} \, 
(\xi_{i + 2} - \xi_{i}) \, \left[
\xi_{j}^2 \, S_{j} + 4 \, \xi_{j + 1}^2 \, S_{j + 1} + \xi_{j+2}^2 \, S_{j+2} \right]
                        \right|.
\label{Wgravpoten} 
\end{equation}

Finally, the angular momentum $J$ is computed by the relation
\begin{equation}
J= \, [J]_\mathrm{cpu} \, \Omega_* \, 
\sum_{j=1(2)}^\mathrm{KRP-2} \, \frac{1}{6} \, 
(\xi_{i + 2} - \xi_{i}) \left[
\xi_{j}^4 \, Q_j + 4 \, \xi_{j + 1}^4 \, Q_{j + 1} + \xi_{j+2}^4 \, Q_{j+2}
                        \right],
\label{Jang} 
\end{equation} 
where the auxiliary array $\{Q_j\}$ is that computed by Eq.~(\ref{QforTJ}).

\section{Numerical Results and Discussion}
\label{rd}
We first compute general-relativistic polytropic models of maximum mass, $M_\mathrm{max}$, in critical rotation with $n=2.9, \, 2.5, \, 2.0, \, 1.5, \, \mathrm{and} \, 1.0$. The case $n=2.9$ represents the softest EOS among those resolved, while the case $n=1.0$ represents the stiffest one.  

A discussion on models of maximum mass can be found in \citep{GS11} (Sec.~4 and references therein); in the present study, we apply the procedure described there for computing the central rest-mass density $\rho_\mathrm{c}^\mathrm{max} = \rho_\mathrm{c}(M_\mathrm{max})$ of a model of maximum mass. Next, we find the central pressure $p_\mathrm{c}^\mathrm{max}$ by Eq.~(\ref{2.5}), and the mass-energy density $E_\mathrm{c}^\mathrm{max}$ by using the relations~(\ref{Erho})--(\ref{secondseq}). For the polytropic constant $K$, we choose the same values with those in \citep{GS11} (Tables 2--6). The gravitation parameter $\sigma_\mathrm{max}$ is then calculated by Eq.~(\ref{2.7}b). 

For decreasing $n$, the values $\sigma_\mathrm{max}$ get increasing; namely, the softest case $n=2.9$ has $\sigma_\mathrm{max} \simeq 0.004$, while the stiffest one $n=1.0$ has $\sigma_\mathrm{max} \simeq 0.3$. Since $\sigma_\mathrm{max}$ is large for $n = 1.0$, we find interesting to study two further models for this case with $\sigma = \sigma_\mathrm{max}/2$ and $\sigma_\mathrm{max}/3$, respectively. The corresponding values $\rho_\mathrm{c}$ are found by writing Eq.~(\ref{2.7}b) in the form    
\begin{equation}
\sigma = \frac{1}{c^2} \, \, \frac{K \, \rho_\mathrm{c}^\Gamma}{\rho_{\mathrm{c}}} =
\frac{1}{c^2} \, K \, \rho_\mathrm{c}^{1/n},  
\label{2.7bb}
\end{equation}
and by solving it for $\rho_\mathrm{c}$,
\begin{equation}
\rho_\mathrm{c} = \left( \frac{c^2}{K} \, \, \sigma \right)^n.
\label{2.7bbb}
\end{equation}

Regarding rotation, we study models of maximum mass in critical rotation, i.e., having angular velocities equal to their Keplerian angular velocities $\Omega_\mathrm{K}$. Newtonian configurations are characterized by an angular velocity $\Omega_\mathrm{max}$ given by
\begin{equation}
\Omega_\mathrm{max} = \sqrt{\frac{G\,M}{R^3}},
\label{Wmax}
\end{equation}
which is the maximum angular velocity, for which mass shedding does not yet occur at the equator. Apparently, $\Omega_\mathrm{max}$   describes the Newtonian balance of centrifugal and gravitational forces. However, it is an overestimated limit for relativistic objects, for which the upper bound is instead the Keplerian angular velocity $\Omega_\mathrm{K}$. If the angular velocity of the configuration is slightly greater than $\Omega_\mathrm{K}$, then mass shedding occurs at the equator. Thus $\Omega_\mathrm{K}$ is the relativistic analog of $\Omega_\mathrm{max}$. Several methods have been developed for the computation of $\Omega_\mathrm{K}$. A discussion on appropriate methods is given in \citep{PG03} (Sec.~3.7). A detailed description of such a method can be found in \citep{BFGM05} (Sec.~IIA). This method, slightly modified, is used in \citep{SG12} for computing $\Omega_\mathrm{K}$ by applying the ``complex-plane strategy in the framework of Hartle's perturbation method'' (HCPS), keeping terms of up to third order in $\Omega$.

In the framework of HAS, $\Omega_\mathrm{K}$ is computed by the procedure described in Sec.~\ref{comp}. In particular, after having computed the critical rotation parameter $\upsilon_\mathrm{c}$, we find $\Omega_\mathrm{K}$ by Eq.~(\ref{Omega*}),
\begin{equation}
\left( \Omega_\mathrm{K} \right)_\mathrm{cpu} = \Omega_*(\upsilon_\mathrm{c}) =
\sqrt{\frac{\upsilon_\mathrm{c}}{2}} \, .
\label{OmegaK*}
\end{equation}
An interesting issue related to $\Omega_\mathrm{K}$ has to do with a remark made by Fahlman \& Anand in \citep{FA71} (Sec.~5; that particular PNA's scheme is of first order in $\sigma$, and of second order in $\sigma \upsilon$ and $\upsilon$). According to this remark, terms of order $\sigma \upsilon$ are generally opposite in sign to the corresponding terms in $\upsilon^2$ and, hence, these second-order terms tend to cancel each other. In the framework of HAS, however, relativistic and rotational effects are assumed decoupled (Sec.~\ref{intro}); hence, terms in $\upsilon^2$ remain without their counterbalancing terms in $\sigma \upsilon$. Therefore, it is of interest to find which values $\Omega_\mathrm{K}$ are closer to respective values computed by an alternative numerical method: the ones derived by keeping only terms in $\upsilon$, or those derived by including terms in $\upsilon^2$.
To compute ``reference values'' for $\Omega_\mathrm{K}$, we use in this study the well-known RNS package \citep{S92} with grid size  $\mathrm{MDIV}\times\mathrm{SDIV}=129 \times 257$, accuracy $a=10^{-6}$ and tolerance $b=10^{-5}$. RNS is an accurate, nonperturbative, iterative method; on the other hand, HAS is a perturbative, noniterative method; so, comparing HAS results with respective RNS results seems to be a decisive test for HAS. 

In Table~\ref{tab:OMGAK} we quote percent differences 
\begin{equation}
\%D(\Omega_\mathrm{K}) = 100 \, [(\Omega_\mathrm{K})_\mathrm{HAS} - (\Omega_{K})_\mathrm{RNS}]/(\Omega_\mathrm{K})_\mathrm{HAS}
\end{equation}
of HAS values relative to RNS values. We find that the first-order HAS values are closer to those of RNS, except for the softest case $n=2.9$, for which the second-order HAS value is closer to that of RNS. The small $\sigma_\mathrm{max}$ of this case permits us to work with surface parameter $\tau_\mathrm{s}=0$; all other cases are resolved with $\tau_\mathrm{s} = \sigma/10$ (see however the remarks regarding the stiffest case $n = 1.0$ in the next paragraph). Accordingly, we quote numerical results of first order in $\upsilon$ for the models with $n \leq 2.5$, and of second order in $\upsilon$ for the model $n=2.9$. 
 
It is worth clarifying here that the boundary condition~(\ref{Fgtt}) holding for $n = 1.0$ induces a shrinking of the configuration, since the derived boundary lies inside the physical boundary, tending to cooperate with the relativistic effects; so, the configuration can sustain a larger $\upsilon_\mathrm{c}$, i.e. a larger $\Omega_\mathrm{K}$, in comparison with that of the case $\tau_\mathrm{s} = 0$. On the other hand, the boundary condition~(\ref{Fltt}) holding for $n = 1.5, \, 2.0, \, \mathrm{and} \, \, 2.5$ induces an expansion of the configuration, since the derived  boundary lies outside the physical boundary, tending to cooperate with the rotational effects; so, the configuration can sustain a smaller $\upsilon_\mathrm{c}$, i.e. a smaller $\Omega_\mathrm{K}$, in comparison with that of the case $\tau_\mathrm{s} = 0$. Both boundary conditions lead to values of $\Omega_\mathrm{K}$ closer to those of RNS. Accordingly, physical characteristics related strongly to rotation (i.e. $\Omega_\mathrm{K}$, $R_\mathrm{e}$, $T$, and $J$) obtain values closer to the ones of RNS. However, regarding the case $n = 1.0$, physical characteristics related strongly to gravitation (i.e. all kinds of mass defined in Sec.~\ref{physchar}, and $W$) obtain values appreciably overestimated with respect to those of RNS due to the intensified shrinking effectes discussed above. Therefore, particularly for the cases $n = 1.0, \, \sigma = \sigma_\mathrm{max}$ (Table~\ref{tab:295}) and $n = 1.0, \, \sigma = \sigma_\mathrm{max}/2$ (Table~\ref{tab:296}), the values of $M$, $M_0$, $M_\mathrm{P}$, and $W$ quoted in the tables are those computed by counterbalancing the additional shrinking effects owing to $\tau_\mathrm{s} > 0$, that is, by putting $\tau_\mathrm{s} = 0$ in the relevant computations.

For brevity, we will drop hereafter the superscript ``max'' from the maximum-mass central densities $\rho_\mathrm{c}$ and $E_\mathrm{c}$.      

Tables~\ref{tab:290}--\ref{tab:295} show numerical results for the physical characteristics discussed in Secs.~\ref{physchar} and \ref{comp2}. As compared to RNS, the  HAS values exhibit very satisfactory accuracy for the softest case $n=2.9$ and also for the soft case $n = 2.5$. In particular, in the case $n=2.9$ (Table~\ref{tab:290}) the larger value $|\%D|_\mathrm{max} \simeq 0.9$ occurs for $R_\mathrm{e}$ and the average percent difference is $|\%D|_\mathrm{avr} \sim 0.5$. Likewise, in the case $n=2.5$ (Table~\ref{tab:291}) the larger value $|\%D|_\mathrm{max} \simeq 1.5$ occurs for $J$ and the average percent difference is $|\%D|_\mathrm{avr} \sim 0.5$. In addition, Table~\ref{tab:291} shows results computed by HCPS (\citep{SG12}, Table~4; the value of $\Omega_\mathrm{K}$ quoted there has been computed by RNS), and their percent differences relative to respective RNS results. Note that the case $n = 2.5$ is the softest one resolved in \citep{SG12}.  

Next, we verify a satisfactory accuracy for the moderately stiff case $n=2.0$ (Table~\ref{tab:292}), where the larger value $|\%D|_\mathrm{max} \simeq 2.5$ occurs for $J$ and the average percent difference is $|\%D|_\mathrm{avr} \sim 1$. Likewise, for the moderately stiff case $n=1.5$ (Table~\ref{tab:293}) we find that the larger value $|\%D|_\mathrm{max} \simeq 5.3$ appears for $|W|$ and the average percent difference is $|\%D|_\mathrm{avr} \sim 2.5$. 

On the other hand, there is a tolerable accuracy, at least concerning $|\%D|_\mathrm{avr}$,  for the stiffest case $n=1.0$. In particular, Table~\ref{tab:295} shows that the larger value $|\%D|_\mathrm{max} \simeq 12$ arises for $T$, while the average percent difference is $|\%D|_\mathrm{avr} \sim 4.5$. In addition, Table~\ref{tab:295} shows results computed by HCPS (\citep{SG12}, Table~1; the value of $\Omega_\mathrm{K}$ quoted there has been computed by RNS), and their percent differences relative to respective RNS results. The case $n = 1.0$ is the stiffest one resolved in \citep{SG12}.  

Second, since for the case $n = 1.0$ the value of $\sigma_\mathrm{max}$ gets large, we find interesting to study two further models having instead $\sigma = \sigma_\mathrm{max}/2$ (Table~\ref{tab:296}) and $\sigma_\mathrm{max}/3$ (Table~\ref{tab:297}). The first model exhibits its larger value $|\%D|_\mathrm{max} \simeq 8.1$ for $T$ and average percent difference $|\%D|_\mathrm{avr} \sim 3.5$. The second model has $|\%D|_\mathrm{max} \simeq 6.2$, occuring for $R_\mathrm{e}$, and $|\%D|_\mathrm{avr} \sim 2.5$. Our results show that both the larger and the average percent differences get decreasing as $\sigma$ decreases; in fact, the model with $\sigma = \sigma_\mathrm{max}/3$ exhibits an accuracy compatible with that of the maximum-mass, critically rotating model $n = 1.5$.  

Third, we also find interesting to study a maximum-mass model with $n=1.0$ in very rapid rotation, having angular velocity 
\begin{equation}
\Omega_* = \Omega_* \left(\upsilon_\mathrm{c}/2 \right) = 
           \sqrt{\frac{1}{2} \, \frac{\upsilon_\mathrm{c}}{2}} =
           \sqrt{\frac{1}{2}} \ \left( \Omega_\mathrm{K} \right)_\mathrm{cpu}
           \simeq 0.7 \, \left( \Omega_\mathrm{K} \right)_\mathrm{cpu}.
\end{equation}
Table~\ref{tab:299} gives the physical characteristics of this model. We find that the larger value $|\%D|_\mathrm{max} \simeq 6.5$ occurs for both $T$ and $J$, while the average percent difference is $|\%D|_\mathrm{avr} \simeq 2.5$. Hence, for this highly relativistic rapidly rotating model, the accuracy achived by HAS is again compatible with that of the maximum-mass, critically rotating model $n = 1.5$.

\section{Concluding Remarks}
Focusing on physical characteristics related strongly to rotation, we remark that HAS computes results, which are close to those of RNS. It is well-known that most perturbative, noniterative methods have great difficulties in computing with satisfactory accuracy quantities like $\Omega_\mathrm{K}$ and $R_\mathrm{e}$. A detailed discussion on this matter can be found in \citep{BFGM05} (Sec.~III). Regarding $R_\mathrm{e}$ in critical rotation (in fact, in mass-shedding limit), Tables~II and III in \citep{BFGM05} quote discrepancies relative to results of nonperturbative methods used in \citep{CST94424} and \citep{BS04} from $\sim 15\%$ to $\sim 25\%$, dependent on the particular models studied (namely, constant-mass and maximum-mass sequences, respectively). In addition, Table~IV in \citep{BFGM05} quotes values of $\Omega_\mathrm{K}$ (in fact, mass-shedding frequencies $\nu_\mathrm{ms}$) with discrepancies, relative to results of \citep{CST94424} and \citep{BS04}, from $\sim 20\%$ to $\sim 25\%$. Furthermore, Tables~\ref{tab:291} and \ref{tab:295} incorporate relevant results of two of the models resolved in \citep{SG12} (Sec.~7, Tables~4 and 1, respectively) by using HCPS: those with $n = 2.5$ and $n = 1.0$, respectively, which represent the softest and stiffest cases studied in \citep{SG12}. The discrepancies regarding $R_\mathrm{e}$ values, relative to RNS, are $\sim 35\%$ and $\sim 30\%$, respectively. In addition, Tables~6,\,7, and 8 in \citep{SG12} show $\Omega_\mathrm{K}$ values (in fact, mass-shedding angular velocities $\Omega_\mathrm{MS}$) for models of constant baryonic mass with $n = 1.0, \, 1.5, \, \mathrm{and} \, \, 2.5$, respectively, computed by HCPS. The discrepancies relative to RNS are from $\sim 17\%$ to $\sim 23\%$.

On the other hand, the results computed by HAS are much closer to those of RNS. In particular, the larger discrepancy concerning $\Omega_\mathrm{K}$ is $\sim 6.5\%$ (Table~\ref{tab:OMGAK}, fifth entry: case $n = 1.0$); while the larger discrepancy concerning $R_\mathrm{e}$ for maximum-mass models is $\sim 5.5\%$ (Table~\ref{tab:293}, fourth entry: case $n = 1.5$). In conclusion, as compared to RNS, HAS is proved to be accurate and reliable for computing models in the extreme regime of maximum mass and critical rotation, from the softest case $n = 2.9$ to the stiffest one $n = 1.0$.

Finally, it should be stressed that HAS is a fast numerical method. In particular, by comparing execution times of HAS and RNS on the same computer, we have verified that HAS is $\sim \! 25$ times faster than RNS for the model $n = 1.0$, $\sim \! 5$ times faster for the model $n = 2.0$, and $\sim \! 3$ times faster for the models $n = 1.5, \, 2.5, \, \mathrm{and} \, \, 2.9$.   
 
\begin{table}
\caption{Percent differences $\%D(\Omega_\mathrm{K})$ of the Keplerian angular velocities $\Omega_\mathrm{K}$ computed by HAS relative to respective values computed by RNS. Labels ``R1'' and ``R2'' denote first-order and second-order results in the rotation parameter $\upsilon$, respectively. The parenthesized signed integers, following numeric values, denote powers of ten.\label{tab:OMGAK}}
\begin{center}
\begin{tabular}{lrrr} 
\hline \hline
$n$  & $\sigma~~~~~~~~~~~~~~~$ & R1~~~~~ & R2~~~~~   \\ 
\hline
$2.9$ & $\sigma_\mathrm{max}=4.41591(-03)$  & $8.114(-01)$  & $5.198(-02)$        \\ 
$2.5$ & $\sigma_\mathrm{max}=2.68066(-02)$  & $1.224(+00)$  & $1.678(+00)$        \\ 
$2.0$ & $\sigma_\mathrm{max}=7.10464(-02)$  & $2.235(+00)$  & $5.473(+00)$        \\ 
$1.5$ & $\sigma_\mathrm{max}=1.50569(-01)$  & $2.988(+00)$  & $3.659(+00)$        \\ 
$1.0$ & $\sigma_\mathrm{max}=3.19773(-01)$  & $-6.403(+00)$  & $-1.114(+01)$      \\ 
$1.0$ & $\sigma_\mathrm{max}/2=1.59887(-01)$  & $-6.400(+00)$  & $-1.104(+01)$    \\ 
$1.0$ & $\sigma_\mathrm{max}/3=1.06591(-01)$  & $-6.035(+00)$  & $-1.046(+01)$    \\
\hline
\end{tabular}
\end{center}
\end{table} 

\begin{table}
\caption{Physical characteristics of a general-relativistic, maximum-mass, critically rotating polytropic model with $n=2.9$, $\rho_\mathrm{c} = 1.481012(-07)$, $E_\mathrm{c} = 1.499978(-07)$, $\sigma_\mathrm{max}=4.41591(-03)$. In all required conversions, we use the value $K = 2.6(+13) \, \mathrm{cgs}$. Columns ``HAS'' and ``RNS'' show results computed by HAS and RNS, respectively. Column ``$\%D$'' shows percent differences $\%D(X) = 100 \, (X_\mathrm{HAS} - X_\mathrm{RNS})/X_\mathrm{HAS}$. All quantities (except for $K$, which is given in cgs, and for the dimensionless ratio in the last entry) are given in polytropic units related to the gravitational units (pu). The parenthesized signed integers, following numeric values, denote powers of ten.\label{tab:290}}
\begin{center}
\begin{tabular}{lrrr} 
\hline \hline
quantity  & HAS~~~~ & RNS~~~~ & $\%D$~~~~ \\ 
\hline
$M$   & $3.323(+00)$ & $3.323(+00)$ & $1.815(-03)$     \\ 
$M_0$ & $3.324(+00)$ & $3.324(+00)$ & $-1.038(-02)$     \\ 
$M_\mathrm{P}$ & $3.348(+00)$ & $3.348(+00)$ & $-1.034(-02)$       \\ 
$R_\mathrm{e}$ & $9.031(+02)$ & $9.114(+02)$ & $-9.232(-01)$      \\
$\Omega_\mathrm{K}$ & $6.639(-05)$ & $6.635(-05)$ & $5.198(-02)$   \\  
$T$ & $2.545(-04)$ & $2.556(-04)$ & $-4.499(-01)$     \\
$|W|$ & $2.540(-02)$ & $2.536(-02)$ & $1.767(-01)$    \\
$J$ & $7.666(+00)$ & $7.705(+00)$ & $-5.021(-01)$     \\
$R_\mathrm{p}/R_\mathrm{e}$ & $6.688(-01)$ & $6.614(-01)$ & $1.100(+00)$ \\
\hline
\end{tabular}
\end{center}
\end{table} 

\begin{table}
\caption{Physical characteristics of a general-relativistic, maximum-mass, critically rotating polytropic model with $n=2.5$, $\rho_\mathrm{c} = 1.176534(-04)$, $E_\mathrm{c} = 1.255381(-04)$, $\sigma_\mathrm{max}=2.68066(-02)$. In all required conversions, we use the value $K = 1.5(+13) \, \mathrm{cgs}$. Column ``HCPS'' shows results computed by HCPS. Column ``$\%D_\mathrm{HCPS}$'' shows percent differences $\%D_\mathrm{HCPS}(X) = 100 \, (X_\mathrm{HCPS} - X_\mathrm{RNS})/X_\mathrm{HCPS}$. Details as in Table~\ref{tab:290}.\label{tab:291}}
\begin{center}
\begin{tabular}{lrrrrr} 
\hline \hline
quantity  & HAS~~~~ & RNS~~~~ & $\%D$~~~~ & HCPS~~~ & $\%D_\mathrm{HCPS}$~~ \\ 
\hline
$M$   & $1.298(+00)$ & $1.298(+00)$ & $-2.278(-02)$ & $1.293(+00)$ & $-3.867(-01)$     \\ 
$M_0$ & $1.302(+00)$ & $1.303(+00)$ & $-6.539(-02)$ & $1.300(+00)$ & $-2.308(-01)$     \\ 
$M_\mathrm{P}$ & $1.351(+00)$ & $1.351(+00)$ & $-6.490(-02)$ & $1.348(+00)$ & $-2.226(-01)$ \\       
$R_\mathrm{e}$ & $5.956(+01)$ & $5.910(+01)$ & $7.795(-01)$ & $4.394(+01)$ & $-3.450(+01)$    \\
$\Omega_\mathrm{K}$ & $2.543(-03)$ & $2.512(-03)$ & $1.224(+00)$                    \\  
$T$ & $8.448(-04)$ & $8.472(-04)$ & $-2.742(-01)$ & $8.235(-04)$ & $-2.878(+00)$    \\
$|W|$ & $5.466(-02)$ & $5.419(-02)$ & $8.658(-01)$ & $5.620(-02)$ & $3.577(+00)$    \\
$J$ & $6.644(-01)$ & $6.745(-01)$ & $-1.517(+00)$ & $6.557(-01)$ & $-2.867(+00)$    \\
$R_\mathrm{p}/R_\mathrm{e}$ & $6.602(-01)$ & $6.533(-01)$ & $1.043(+00)$ \\
\hline
\end{tabular} 
\end{center}
\end{table}

\begin{table}
\caption{Physical characteristics of a general-relativistic, maximum-mass, critically rotating polytropic model with $n=2.0$, $\rho_\mathrm{c} = 5.047591(-03)$, $E_\mathrm{c} = 5.764817(-03)$, $\sigma_\mathrm{max}=7.10464(-02)$. In all required conversions, we use the value $K = 1.0(+12) \, \mathrm{cgs}$. Details as in Table~\ref{tab:290}.\label{tab:292}}
\begin{center}
\begin{tabular}{lrrr} 
\hline \hline
quantity  & HAS~~~~ & RNS~~~~ & $\%D$~~~~ \\ 
\hline
$M$   & $5.502(-01)$ & $5.494(-01)$ & $1.428(-01)$     \\ 
$M_0$ & $5.598(-01)$ & $5.592(-01)$ & $1.080(-01)$     \\ 
$M_\mathrm{P}$ & $6.008(-01)$ & $6.002(-01)$ & $1.087(-01)$       \\ 
$R_\mathrm{e}$ & $1.008(+01)$ & $9.903(+00)$ & $1.791(+00)$       \\ 
$\Omega_\mathrm{K}$ & $2.436(-02)$ & $2.381(-02)$ & $2.235(+00)$      \\
$T$ & $1.401(-03)$ & $1.404(-03)$ & $-2.461(-01)$   \\  
$|W|$ & $5.332(-02)$ & $5.216(-02)$ & $2.179(+00)$   \\
$J$ & $1.150(-01)$ & $1.179(-01)$ & $-2.537(+00)$     \\
$R_\mathrm{p}/R_\mathrm{e}$ & $6.565(-01)$ & $6.383(-01)$ & $2.769(+00)$ \\
\hline
\end{tabular}
\end{center}
\end{table}

\begin{table}
\caption{Physical characteristics of a general-relativistic, maximum-mass, critically rotating polytropic model with $n=1.5$, $\rho_\mathrm{c} = 5.842562(-02)$, $E_\mathrm{c} = 7.162125(-02)$, $\sigma_\mathrm{max}=1.50569(-01)$. In all required conversions, we use the value $K = 5.3802(+09) \, \mathrm{cgs}$. Details as in Table~\ref{tab:290}.\label{tab:293}}
\begin{center}
\begin{tabular}{lrrr} 
\hline \hline
quantity  & HAS~~~~ & RNS~~~~ & $\%D$~~~~ \\ 
\hline
$M$   & $2.950(-01)$ & $2.905(-01)$ & $1.545(+00)$     \\ 
$M_0$ & $3.094(-01)$ & $3.040(-01)$ & $1.742(+00)$     \\ 
$M_\mathrm{P}$ & $3.420(-01)$ & $3.363(-01)$ & $1.659(+00)$       \\ 
$R_\mathrm{e}$ & $2.902(+00)$ & $2.748(+00)$ & $5.307(+00)$       \\ 
$\Omega_\mathrm{K}$ & $1.217(-01)$ & $1.180(-01)$ & $2.988(+00)$      \\
$T$ & $2.329(-03)$ & $2.235(-03)$ & $4.044(+00)$   \\  
$|W|$ & $5.078(-02)$ & $4.807(-02)$ & $5.333(+00)$   \\
$J$ & $3.828(-02)$ & $3.786(-02)$ & $1.089(+00)$     \\
$R_\mathrm{p}/R_\mathrm{e}$ & $6.118(-01)$ & $6.161(-01)$ & $-7.037(-01)$ \\
\hline
\end{tabular} 

\end{center}
\end{table}

\begin{table}
\caption{Physical characteristics of a general-relativistic, maximum-mass, critically rotating polytropic model with $n=1.0$, $\rho_\mathrm{c} = 3.197730(-01)$, $E_\mathrm{c} = 4.220278(-01)$, $\sigma_\mathrm{max}=3.19773(-01)$. In all required conversions, we use the value $K = 1.0(+05) \, \mathrm{cgs}$. Details as in Table~\ref{tab:291}.\label{tab:295}}
\begin{center}
\begin{tabular}{lrrrrr} 
\hline \hline
quantity  & HAS~~~~ & RNS~~~~ & $\%D$~~~~ & HCPS~~~ & $\%D_\mathrm{HCPS}$~~ \\ 
\hline
$M$   & $1.844(-01)$ & $1.876(-01)$ & $-1.698(+00)$ & $1.789(-01)$ & $-4.863(+00)$    \\ 
$M_0$ & $2.040(-01)$ & $2.061(-01)$ & $-1.014(+00)$ & $1.965(-01)$ & $-4.886(+00)$   \\ 
$M_\mathrm{P}$ & $2.306(-01)$ & $2.332(-01)$ & $-1.111(+00)$ & $2.230(-01)$ & $-4.574(+00)$       \\ 
$R_\mathrm{e}$ & $1.073(+00)$ & $1.032(+00)$ & $3.839(+00)$ & $7.928(-01)$ & $-3.017(+01)$      \\ 
$\Omega_\mathrm{K}$ & $3.835(-01)$ & $4.080(-01)$ & $-6.403(+00)$      \\
$T$ & $3.582(-03)$ & $4.011(-03)$ & $-1.196(+01)$ & $3.471(-03)$ & $-1.556(+01)$  \\  
$|W|$ & $5.086(-02)$ & $4.960(-02)$ & $2.466(+00)$ & $4.753(-02)$ & $-4.355(+00)$   \\
$J$ & $1.868(-02)$ & $1.966(-02)$ & $-5.222(+00)$ & $1.702(-02)$ & $-1.551(+01)$     \\
$R_\mathrm{p}/R_\mathrm{e}$ & $6.082(-01)$ & $5.863(-01)$ & $3.597(+00)$ \\
\hline
\end{tabular}
\end{center}
\end{table}

\begin{table}
\caption{Physical characteristics of a general-relativistic, critically rotating polytropic model with $n=1.0$, $\rho_\mathrm{c} = 1.598870(-01)$, $E_\mathrm{c} = 1.854509(-01)$, $\sigma = \frac{1}{2}\sigma_\mathrm{max}(n=1.0) = 1.59887(-01)$. In all required conversions, we use the value $K = 1.0(+05) \, \mathrm{cgs}$. Details as in Table~\ref{tab:290}.\label{tab:296}}
\begin{center}
\begin{tabular}{lrrr} 
\hline \hline
quantity  & HAS~~~~ & RNS~~~~ & $\%D$~~~~ \\ 
\hline
$M$   & $1.780(-01)$ & $1.790(-01)$ & $-5.491(-01)$     \\ 
$M_0$ & $1.940(-01)$ & $1.947(-01)$ & $-3.477(-01)$     \\ 
$M_\mathrm{P}$ & $2.074(-01)$ & $2.083(-01)$ & $-4.400(-01)$       \\ 
$R_\mathrm{e}$ & $1.372(+00)$ & $1.274(+00)$ & $7.114(+00)$       \\ 
$\Omega_\mathrm{K}$ & $2.772(-01)$ & $2.949(-01)$ & $-6.400(+00)$      \\
$T$ & $2.681(-03)$ & $2.898(-03)$ & $-8.115(+00)$   \\  
$|W|$ & $3.278(-02)$ & $3.223(-02)$ & $1.685(+00)$   \\
$J$ & $1.934(-02)$ & $1.965(-02)$ & $-1.612(+00)$     \\
$R_\mathrm{p}/R_\mathrm{e}$ & $5.751(-01)$ & $5.784(-01)$ & $-5.652(-01)$ \\
\hline
\end{tabular}
\end{center}
\end{table}

\begin{table}
\caption{Physical characteristics of a general-relativistic, critically rotating polytropic model with $n=1.0$, $\rho_\mathrm{c} = 1.065910(-01)$, $E_\mathrm{c} = 1.179526(-01)$, $\sigma = \frac{1}{3} \sigma_\mathrm{max}(n=1.0) = 1.06591(-01)$. In all required conversions, we use the value $K = 1.0(+05) \, \mathrm{cgs}$. Details as in Table~\ref{tab:290}.\label{tab:297}}
\begin{center}
\begin{tabular}{lrrr} 
\hline \hline
quantity  & HAS~~~~ & RNS~~~~ & $\%D$~~~~ \\ 
\hline
$M$   & $1.608(-01)$ & $1.599(-01)$ & $5.632(-01)$     \\ 
$M_0$ & $1.725(-01)$ & $1.715(-01)$ & $5.953(-01)$     \\ 
$M_\mathrm{P}$ & $1.806(-01)$ & $1.797(-01)$ & $5.137(-01)$       \\ 
$R_\mathrm{e}$ & $1.506(+00)$ & $1.413(+00)$ & $6.175(+00)$       \\ 
$\Omega_\mathrm{K}$ &  $2.274(-01)$ & $2.411(-01)$ & $-6.035(+00)$      \\
$T$ & $2.046(-03)$ & $2.042(-03)$ & $1.705(-01)$   \\  
$|W|$ & $2.236(-02)$ & $2.185(-02)$ & $-1.514(+00)$   \\
$J$ & $1.800(-02)$ & $1.694(-02)$ & $5.852(+00)$     \\
$R_\mathrm{p}/R_\mathrm{e}$ & $5.635(-01)$ & $5.739(-01)$ & $-1.843(+00)$ \\
\hline
\end{tabular}
\end{center}
\end{table}

\begin{table}
\caption{Physical characteristics of a general-relativistic, maximum-mass polytropic model rotating with rotation parameter $\upsilon = \upsilon_\mathrm{c}/2$, i.e. $\Omega = \sqrt{1/2} \, \, \Omega_\mathrm{K}$; $n=1.0$, $\rho_\mathrm{c} = 3.197730(-01)$, $E_\mathrm{c} = 4.220278(-01)$, $\sigma_\mathrm{max}=3.19773(-01)$. Details as in Table~\ref{tab:295}.\label{tab:299}}
\begin{center}
\begin{tabular}{lrrr} 
\hline \hline
quantity  & HAS~~~~ & RNS~~~~ & $\%D$~~~~ \\ 
\hline
$M$   & $1.707(-01)$ & $1.694(-01)$ & $7.168(-01)$     \\ 
$M_0$ & $1.875(-01)$ & $1.861(-01)$ & $7.506(-01)$    \\ 
$M_\mathrm{P}$ & $2.130(-01)$ & $2.116(-01)$ & $6.625(-01)$       \\ 
$R_\mathrm{e}$ & $7.712(-01)$ & $7.968(-01)$ & $-3.318(+00)$       \\ 
$\Omega$ &  $2.342(-01)$ & $2.342(-01)$ & $0.000(+00)$      \\
$T$ & $9.099(-04)$ & $9.691(-04)$ & $-6.504(+00)$   \\  
$|W|$ & $4.545(-02)$ & $4.315(-02)$ & $5.057(+00)$   \\
$J$ & $7.771(-03)$ & $8.276(-03)$ & $-6.503(+00)$     \\
$R_\mathrm{p}/R_\mathrm{e}$ & $8.983(-01)$ & $9.043(-01)$ & $-6.707(-01)$ \\
\hline
\end{tabular} 
\end{center}
\end{table}

\end{document}